\documentclass[aps,prd,preprint,superscriptaddress,showpacs,preprintnumbers]{revtex4-1}
\usepackage[colorlinks=true, pdfstartview=FitV, linkcolor=red, citecolor=blue, urlcolor=black, pdftitle={},pdfauthor={Tomoya Hayata},pdfsubject={}, pdfkeywords={}]{hyperref}
\usepackage[pdftex]{graphicx}
\usepackage{setspace,bm}
\usepackage{latexsym,amssymb,amsmath,mathrsfs}
\usepackage{color}
\graphicspath{{./figure/}}

\newcommand{\be}{\begin{equation}}      
\newcommand{\ee}{\end{equation}}      
\newcommand{\bea}{\begin{eqnarray}}      
\newcommand{\eea}{\end{eqnarray}}

\newcommand{\tr}{\mathrm{tr}}
\newcommand{\im}{\mathrm{i}}
\newcommand{\pf}{\mathop{\mathrm{Pf}}}
\renewcommand{\S}{{\mathrm{S}}}
\newcommand{\AS}{\mathrm{AS}}

\begin{document}
\title{Classification of sign-problem-free relativistic fermions on the basis of the Majorana positivity}

\author{Tomoya Hayata}
\affiliation{Department of Physics, Chuo University, Tokyo, 112-8551, Japan}
\author{Arata Yamamoto}
\affiliation{Department of Physics, The University of Tokyo, Tokyo 113-0031, Japan}

\date{\today}

\begin{abstract}
We classify the sign-problem-free relativistic fermion actions on the basis of the Majorana representation.
In the Majorana representation, the sign-problem-free condition is given by the semi-positivity of a Pfaffian.
We show that the known sign-problem-free actions of the Dirac fermions, which are usually understood from the semi-positivity of a determinant, e.g., the action of quantum chromodynamics with nonzero chiral chemical potential or nonzero isospin chemical potential, can also be understood from the semi-positivity of a Pfaffian.
We also derive new classes of the sign-problem-free relativistic fermion actions with Majorana-type source terms.
\end{abstract}

\maketitle

\section{Introduction}

The quantum Monte Carlo method is a powerful computational scheme in modern physics from particle physics to condensed matter physics.
In the Monte Carlo method, the semi-positivity of a weight factor is crucial.
When the semi-positivity is lost, the method breaks down due to the sign fluctuation.
This is called the sign problem.
The sign problem is frequently induced by fermions.
A famous example of the fermion sign problem is quantum chromodynamics (QCD) with nonzero baryon chemical potential \cite{[{For reviews, }]Philipsen:2005mj,*Stephanov:2007fk,*deForcrand:2010ys,*Aarts:2013lcm}.
Although many challenges have been done for a long time, the problem has not yet been solved.
In recent years, the novel attempts to evaluate complex integral, such as the complex Langevin method and the Lefschetz thimble, are intensively discussed \cite{[{For reviews, }]Sexty:2014dxa,*Scorzato:2015qts}.
However, their applications are still primitive and limited because of several difficulties \cite{Aarts:2013uxa,Nishimura:2015pba,Tanizaki:2015rda,Hayata:2015lzj}.
The fermion sign problem will remain as an unsolved challenge for the near future.

Under such circumstances, it is important to study sign-problem-free fermions, i.e., fermions with a semi-positive weight factor.
A well-known example is the two-flavor Dirac fermion with isospin chemical potential.
The conventional proof for its semi-positivity is the use of the $\gamma_5$-hermiticity and double degeneracy of the Dirac determinant.
However, this proof is specific to this case and not applicable to general cases.
Although many sign-problem-free fermions might be hiding, we do not know a systematic way to seek them.

It was recently proposed that the concept of the Majorana representation is useful to find sign-problem-free fermions \cite{Li:2014tla,2016PhRvL.116y0601W,Li:2016gte}.
In the Majorana representation, we can systematically prove the semi-positivity even if a determinant is not doubly degenerate or even if a weight factor is not given by a determinant.
This opens up the possibility to discover new classes of sign-problem-free fermions.
Several new sign-problem-free fermion models were actually found in condensed matter physics \cite{2016PhRvL.116y0601W,Li:2016gte}.

In this paper, we discuss sign-problem-free relativistic fermions in four dimensions on the basis of the Majorana representation.
In Sec.~\ref{sec:maj}, we introduce the Majorana positivity condition, that is, a sufficient condition to show the semi-positivity of a Pfaffian.
By using the Majorana positivity condition, we discuss sign-problem-free fermions in one-flavor case in Sec.~\ref{sec:one}, and in two-flavor case in Sec.~\ref{sec:two}.
We list several comments in Sec.~\ref{sec:mis}.
Finally we summarize this paper in Sec.~\ref{sec:sum}.
The derivation of the Majorana positivity condition, and the definition of the Euclidean gamma matrices are summarized in Appendices~\ref{sec:appA}, and~\ref{sec:appB}.

\section{Majorana positivity} 
\label{sec:maj}
Let the Euclidean Majorana action 
\be
S =\frac{1}{2}\int d^4x \Psi^\top P \Psi ,
\label{eq:SM}
\ee
where $\Psi$ is the Majorana fermion field.
Because of the Grassmannian nature of the Majorana fermion, 
only antisymmetric components of $P$ contribute to the Grassmann integration, and we set $P$ to an antisymmetric matrix without loss of generality. 
Then the generating functional $Z$ is given as the real Grassmann integral, and is expressed by the Pfaffian $\pf (P)$ as
\be
Z = \int D\Psi \  e^{-S} = \pf (P)  .
\label{eq:ZM}
\ee
Therefore the standard numerical simulations on the basis of the Monte Carlo sampling do not suffer from the fermionic sign problem when $\pf(P)$ is semi-positive.  
Let us consider the case that the antisymmetric $2N\times 2N$ matrix $P$ is given by a block matrix form
\be
P=
\begin{pmatrix}
P_1 & \im P_2 \\
-\im P_2^\top & P_3
\end{pmatrix}
,
\ee
where $P_i$ are $N\times N$ complex matrices, and $N$ is an even number.
Because of the antisymmetry of $P$, $P_1^\top=-P_1$ and $P_3^\top=-P_3$.
$\pf(P)$ is positive semidefinite if
\begin{description}
 \item[condition 1] $P_2$ is semi-positive,
 \item[condition 2] $P_3 = - P_1^\dagger$ and $P_2 = P_2^\dagger$
\end{description}
are satisfied.
The proof is given in Appendix~\ref{sec:appA}.
This is a sufficient condition for the semi-positivity of a Pfaffian.
We call it the Majorana positivity condition, following Ref.~\cite{2016PhRvL.116y0601W}.
In general, the matrix $P$ has the ambiguity of the basis transformation, which keeps the Pfaffian invariant.
The Pfaffian is semi-positive as long as $P$ satisfies the conditions $1$ and $2$ in one basis, even if it does not satisfy them in other bases.

Two remarks are in order: (I) The semi-positivity is valid regardless of the spin representation of fermion fields, so that the condition can generally be applied to real Grassmann fields, not only relativistic Majorana fermions (Majorana spinors).
(II) The Majorana positivity condition was derived in the Hamiltonian formalism, where the generating functional is given by the functional trace $\tr(e^{-\beta H})$ \cite{2016PhRvL.116y0601W,2016CMaPh.346.1021J}.
Here, we discuss the condition in the Lagrangian formalism, where the generating functional is given by the Pfaffian $\pf(P)$.
The conditions for $P$ to assure the semi-positivity of $\pf(P)$ are slightly different from those for the Hamiltonian $H$ to assure the semi-positivity of $\tr(e^{-\beta H})$.
We need to assume that the block matrices $P_i$ are even-dimensional matrices, which is not assumed in Refs.~\cite{2016PhRvL.116y0601W,2016CMaPh.346.1021J}.
Because of the spinor structure of relativistic fermions, this assumption always holds in the following, so that it is enough to study the constraints to the Dirac operators by the conditions $1$ and $2$.
For details, see Appendix~\ref{sec:appA}.

This argument can be applied to the Dirac fermion integral.
The Dirac field $\psi$ and its conjugate field $\bar{\psi}$ are expressed by using two Majorana fields $\Psi^{(1)}$ and $\Psi^{(2)}$ \cite{Montvay:2001aj} as
\begin{eqnarray}
 \psi &=& \frac{1}{\sqrt{2}} (\Psi^{(1)} + \im \Psi^{(2)} )
 ,
\\
 \psi_c = C\bar{\psi}^\top &=& \frac{1}{\sqrt{2}} (\Psi^{(1)} - \im \Psi^{(2)} )
,
\end{eqnarray}
and
\begin{eqnarray}
 \Psi^{(1)} &=& \frac{1}{\sqrt{2}} (\psi + \psi_c)
,
\\
 \Psi^{(2)} &=& \frac{\im}{\sqrt{2}} (-\psi + \psi_c)
.
\end{eqnarray}
With the two-component Majorana field $\Psi^\top = (\Psi^{(1)\top},\Psi^{(2)\top})$, the Dirac action becomes the Majorana action \eqref{eq:SM}, and the generating functional becomes
\be
\begin{split}
Z &= \int D\bar{\psi} D \psi \ e^{-S[\bar{\psi}, \psi]} 
\\
&= \int D\Psi e^{-S[\Psi]} 
= \pf (P)  
.
\end{split}
\ee
The generating functional is semi-positive if $P$ satisfies the conditions 1 and 2.

\section{One flavor} 
\label{sec:one}

We consider the one-flavor Dirac fermion action
\begin{equation}
\begin{split}
S 
&= \int d^4x \left[ \bar{\psi} D \psi + \bar{\psi}_c d \psi + \bar{\psi} d' \psi_c \right]
\\
&= \int d^4x \left[ \psi^\top_c CD \psi + \psi^\top C d \psi + \psi^\top_c C d' \psi_c \right]
,
\end{split}
\end{equation}
with the Dirac operators
\begin{eqnarray}
 D &=& M + \gamma_5 M_5 + \gamma_\mu D_\mu  +\gamma_\mu\gamma_5 D_{5\mu} + \gamma_\mu\gamma_\nu \Sigma_{\mu\nu}
,
\label{eq:fermion_matrix1}
\\
 d &=& m + \gamma_5 m_5 + \gamma_\mu d_\mu  +\gamma_\mu\gamma_5 d_{5\mu} + \gamma_\mu\gamma_\nu \sigma_{\mu\nu}
,
\label{eq:fermion_matrix2}
\\
 d'&=& m' + \gamma_5 m'_5 + \gamma_\mu d'_\mu +\gamma_\mu\gamma_5 d'_{5\mu} + \gamma_\mu\gamma_\nu \sigma'_{\mu\nu}
,
\label{eq:fermion_matrix3}
\end{eqnarray}
and the charge conjugation matrix $C = \gamma_2 \gamma_4$.
All parameters such as $M$, $D_\mu$, and $\Sigma_{\mu\nu}$ are complex matrices. 
These matrices act on the product space of space-time coordinates and internal gauge group.
The tensor parameters $\Sigma_{\mu\nu}$, $\sigma_{\mu\nu}$, and $\sigma^\prime_{\mu\nu}$ are traceless with respect to the Lorentz indices, e.g., $\Sigma_{\mu\mu}=0$.
To convert the matrix to the form \eqref{eq:positive}, we perform the transformation $\psi \to C_L \psi$ and $\psi_c \to -\gamma_4 C_L \psi_c$, with $C_L={\rm diag}(1,-\im \sigma_2)$ in the chiral basis.
(See Appendix~\ref{sec:appB} for the explicit form of the gamma matrices in chiral basis.)
Then the action becomes
\begin{equation}
S \to \int d^4x \left[ \psi^\top_c H \psi + \psi^\top h \psi + \psi^\top_c h' \psi_c \right]
,
\end{equation}
with
\bea
H &=& - C_L^\top \gamma_4^\top CD C_L \equiv
 \begin{pmatrix}
 H_{RR} & H_{RL} \\
 H_{LR} & H_{LL}
 \end{pmatrix}
,
\label{eq:oneh1}
\\
h &=& C_L^\top C d C_L \equiv
 \begin{pmatrix}
 h_{RR} & h_{RL} \\
 h_{LR} & h_{LL}
 \end{pmatrix}
,
\label{eq:oneh2}
\\
h' &=& C_L^\top \gamma_4^\top C d' \gamma_4 C_L \equiv
 \begin{pmatrix}
 h_{RR}' & h_{RL}' \\
 h_{LR}' & h_{LL}'
 \end{pmatrix}
.
\label{eq:oneh3}
\eea
(The matrix elements are explicitly shown in Appendix~\ref{sec:appB}.)
Because of the Grassmannian nature of $\psi$ and $\psi_c$, the symmetric parts of $h$ and $h'$ do not contribute to the integral.
Thus, we set $h^\S = h'^\S =0$, i.e.,
\begin{eqnarray}
& m^\AS = m^\AS_5 = d_\mu^\S = d_{5\mu}^\AS = \sigma^\S_{\mu\nu} = 0, &
\label{eq:zero1}
\\
& m'^\AS = m'^\AS_5 = d'^\S_\mu = d'^\AS_{5\mu} = \sigma'^\S_{\mu\nu} = 0, &
\label{eq:zero2}
\end{eqnarray}
without loss of generality.
The symmetric and anti-symmetric parts of a complex matrix $O$ are defined by $O^{\rm S}=(O+O^\top)/2$ and $O^{\rm AS}=(O-O^\top)/2$, respectively.

Introducing the Majorana field $\Psi^\top= (\Psi^{(1)\top},\Psi^{(2)\top}) = (R^{(1)\top},L^{(1)\top},R^{(2)\top},L^{(2)\top})$, 
and changing the basis as $(R^{(1)\top},L^{(1)\top},R^{(2)\top},L^{(2)\top}) \to (R^{(1)\top},R^{(2)\top},L^{(2)\top},-L^{(1)\top})$, 
we obtain
\begin{equation}
S = \int d^4x \frac{1}{2} \Psi^\top 
 \begin{pmatrix}
 P_1 & \im P_2
 \\
 -\im P_2^\top & P_3
 \end{pmatrix}
\Psi
,
\end{equation}
with
\be
P_1 =
 \begin{pmatrix}
      H_{RR}^\AS +h_{RR}^\AS +h_{RR}'^\AS  & \im( H_{RR}^\S  +h_{RR}^\AS -h_{RR}'^\AS)
 \\
 \im(-H_{RR}^\S  +h_{RR}^\AS -h_{RR}'^\AS) &      H_{RR}^\AS -h_{RR}^\AS -h_{RR}'^\AS 
 \end{pmatrix}
,
\ee
\be
P_2 =
 \begin{pmatrix}
      H_{RL}^\S  +h_{RL}^\AS -h_{RL}'^\AS & \im( H_{RL}^\AS +h_{RL}^\AS +h_{RL}'^\AS)
 \\
 \im(-H_{RL}^\AS +h_{RL}^\AS +h_{RL}'^\AS)&      H_{RL}^\S  -h_{RL}^\AS +h_{RL}'^\AS
 \end{pmatrix}
,
\ee
\be
P_3 =
 \begin{pmatrix}
      H_{LL}^\AS -h_{LL}^\AS -h_{LL}'^\AS  & \im( H_{LL}^\S  -h_{LL}^\AS +h_{LL}'^\AS) 
 \\
 \im(-H_{LL}^\S  -h_{LL}^\AS +h_{LL}'^\AS) &      H_{LL}^\AS +h_{LL}^\AS +h_{LL}'^\AS  
 \end{pmatrix}
,
\ee
where $H_{RL}^\S = (H_{RL}+H_{LR}^\top)/2$ and so on.
The block matrix $P_2$ must be semi-positive to satisfy the condition 1.
Since the matrix elements of $P_2$ are given by 
$H_{RL}^\S$, $H_{RL}^\AS$, $h_{RL}^\AS$, and $h_{RL}'^\AS$, they give constraints on $M$, $M_5$, $\Sigma_{\mu\nu}$, $d_\mu$, $d'_\mu$, $d_{5\mu}$, and $d'_{5\mu}$.
As for the condition 2, $P_3 = -P_1^\dagger$ is satisfied when 
\begin{eqnarray}
&
D_\mu^\dagger = - D_\mu , \ 
D_{5\mu}^\dagger = D_{5\mu} ,\
&
\label{eq:one1}
\\
&
m^\S = - m'^{\S*} ,\
m^\S_5 = - m'^{\S*}_5, \ 
\sigma^\AS_{\mu\nu} = - \sigma'^{\AS*}_{\mu\nu} ,
&
\label{eq:one2}
\end{eqnarray}
and $P_2 = P_2^\dagger$ is satisfied when
\begin{eqnarray}
&
M^\dagger = M, \ 
M_5^\dagger = M_5 ,\
\Sigma_{\mu\nu}^\dagger = - \Sigma_{\mu\nu} ,\
&
\label{eq:one3}
\\
&
d_\mu^\AS = - d_\mu'^{\AS*} , \ 
d_{5\mu}^\S = - d_{5\mu}'^{\S*}
.
&
\label{eq:one4}
\end{eqnarray}
Equations \eqref{eq:one1}--\eqref{eq:one4}, and the semi-positivity of $P_2$ guarantee the semi-positivity of the Pfaffian.
The result is summarized in Table \ref{tab1}.

\begin{table}[ht]
\centering
\begin{tabular}{|c|c|c|c|}
\hline
 term & condition 1 & condition 2 & examples
\\ \hline \hline
 $\bar{\psi}M\psi$ & \checkmark & $M=M^\dagger$ & Dirac mass
\\ \hline
 $\bar{\psi} \gamma_5 M_5 \psi$ & \checkmark & $M_5=M_5^\dagger$ &
\\ \hline
 & & & gauge field
\\
 $\bar{\psi}\gamma_\mu D_\mu \psi$ & & $D_\mu=-D_\mu^\dagger$ & imaginary chemical potential \cite{Hart:2000ef,deForcrand:2002hgr,*deForcrand:2003vyj,*deForcrand:2006pv,*deForcrand:2008vr,*deForcrand:2010he,D'Elia:2002gd,*D'Elia:2004at,Giudice:2004se,Chen:2004tb,Azcoiti:2004ri,Wu:2006su,Cea:2006yd,*Cea:2007vt,*Cea:2010md,D'Elia:2007ke,Conradi:2007be,D'Elia:2009tm,*D'Elia:2009qz,Cea:2009ba,Bonati:2010gi,Nagata:2011yf,Cea:2012ev,Wu:2013bfa,*Wu:2014lsa,Cea:2014xva,Philipsen:2014rpa,*Philipsen:2016hkv,Yamamoto:2014vda,Takahashi:2014rta,Bonati:2014kpa,Makiyama:2015uwa}
\\
 & & & imaginary orbit-rotation coupling \cite{Yamamoto:2013zwa}
\\ \hline
 & & & imaginary axial gauge field
\\
 $\bar{\psi}\gamma_\mu\gamma_5 D_{5\mu} \psi$ & & $D_{5\mu}=D_{5\mu}^\dagger$ &
 chiral chemical potential \cite{Yamamoto:2011gk,*Yamamoto:2011ks,Buividovich:2013hza,Braguta:2015zta,Braguta:2015owi}
\\
 & & & imaginary spin-rotation coupling \cite{Yamamoto:2013zwa}
\\ \hline
 $\bar{\psi} \gamma_\mu\gamma_\nu \Sigma_{\mu\nu} \psi$ & \checkmark & $\Sigma_{\mu\nu} = - \Sigma_{\mu\nu}^\dagger$ &
\\ \hline
 $\bar{\psi}_c m^\S \psi + \bar{\psi} m'^\S \psi_c$ & & $m^\S = - m'^{\S*}$ & Majorana mass
\\ \hline
 $\bar{\psi}_c \gamma_5 m_5^\S \psi + \bar{\psi} \gamma_5 m_5'^\S \psi_c$ & & $m^\S_5 = - m'^{\S*}_5$ &
\\ \hline
 $\bar{\psi}_c \gamma_\mu d_\mu^\AS \psi + \bar{\psi} \gamma_\mu d_\mu'^\AS \psi_c$ & \checkmark & $d_\mu^\AS = - d_\mu'^{\AS*}$ &
\\ \hline
 $\bar{\psi}_c \gamma_\mu \gamma_5 d_{5\mu}^\S \psi + \bar{\psi} \gamma_\mu \gamma_5 d_{5\mu}'^\S \psi_c$ & \checkmark & $d_{5\mu}^\S = - d_{5\mu}'^{\S*}$ &
\\ \hline
 $\bar{\psi}_c \gamma_\mu \gamma_\nu \sigma_{\mu\nu}^\AS \psi + \bar{\psi} \gamma_\mu \gamma_\nu \sigma_{\mu\nu}'^\AS \psi_c$ & & $\sigma^\AS_{\mu\nu} = - \sigma'^{\AS*}_{\mu\nu}$  &
\\ \hline
\end{tabular}
\caption{\label{tab1}
Summary table of sign-problem-free terms of the one-flavor Dirac fermion.
In the column of condition 1, the checkmark stands for the constraint by the semi-positivity of $P_2$.
}
\end{table}

For example, the standard QCD Dirac operator is
\begin{equation}
 D = \gamma_\mu D_\mu + M,
\quad d = d' = 0 ,
\end{equation}
with $D_\mu = \partial_\mu + \im A_\mu^a T^a = -D_\mu^\dagger$, and real positive $M \in {\bf R}^+$.
It satisfies Eqs.\eqref{eq:one1}--\eqref{eq:one4}, and 
\begin{equation}
 P_2 =
 \begin{pmatrix}
 M & 0
 \\
 0 & M
 \end{pmatrix}
\end{equation}
is positive definite.
Thus it is sign-problem free.
On the other hand, the lattice Wilson-Dirac operator does not satisfy the condition 1, because the Wilson term $M$ is not semi-positive.
This is consistent with the fact that the one-flavor Wilson fermion has the sign problem.
Other known sign-problem-free terms, such as chiral chemical potential, are also explained by this Majorana positivity argument, as shown in Table \ref{tab1}.
In addition, we found new sign-problem-free terms including the Majorana-type terms.
The Majorana-type terms explicitly break gauge symmetry, i.e., particle number conservation, and are used for the source terms of superconductivity.
For example, we can add real $d_{54},d'_{54} \in {\bf R}$ to the QCD Dirac operator.
Equation \eqref{eq:one4} is satisfied when $d_{54} = -d'_{54}$, and
\begin{equation}
 P_2 =
 \begin{pmatrix}
 M +d_{54}-d'_{54} & 0
 \\
 0 & M-d_{54}+d'_{54}
 \end{pmatrix}
\end{equation}
is semi-positive when $ -M \le d_{54}-d'_{54} \le M$.
Thus, the term $\bar{\psi}_c \gamma_4 \gamma_5 d_{54} \psi + \bar{\psi} \gamma_4 \gamma_5 d_{54}' \psi_c$ is sign-problem free when $ -M/2 \le d_{54}= -d'_{54} \le M/2$.

We considered the QCD-type Dirac operator, in which $D_\mu$ is not semi-positive and the Dirac mass $M$ is nonzero.
For this reason, we put $D_\mu$ into $P_1$ and $P_3$, and $M$ into the diagonal components of $P_2$ by the basis transformations.
Otherwise, the condition 1 is not satisfied.
For other types of the Dirac operator, we need to change the basis.
The sign-problem-free terms and their conditions will change.
For example, when $M=0$, pure imaginary $M_5$ ($\im M_5 \in {\bf R}$) becomes sign-problem free, and Eq.~\eqref{eq:one2} becomes $m^\S = m'^{\S*}$, $m^\S_5 = m'^{\S*}_5$, $\sigma^\AS_{\mu\nu} = \sigma'^{\AS*}_{\mu\nu}$.
This can be easily understood from the chiral rotation of the results of positive $M$ given above.

\section{Two flavors}
\label{sec:two}

We consider the two-flavor Dirac fermion action,
\bea
S&=&\int d^4x \bar{\psi}D\psi+ \psi^\top Cd \psi+ \psi_c^\top Cd^{\prime} \psi_c
\nonumber
\\
&=&\int d^4x \psi_c^\top H\psi+ \psi^\top h \psi+ \psi_c^\top h^\prime \psi_c ,
\label{eq:fermion_action}
\eea
where $\psi$ is the two-flavor Dirac field.
The Dirac operators are given by Eqs.~\eqref{eq:fermion_matrix1}, \eqref{eq:fermion_matrix2}, and \eqref{eq:fermion_matrix3}, and the corresponding matrices are given by $H=C D$, $h=Cd$, and $h^\prime=Cd^{\prime}$.
Now the parameters act on the product space of space-time coordinates, internal gauge group, and flavors.
Because of the Grassmannian nature of $\psi$ and $\psi_c$, we set Eqs.~\eqref{eq:zero1} and ~\eqref{eq:zero2}.
By using the two-flavor Majorana fermions, $\Psi^\top=(u^{(1)\top},u^{(2)\top},d^{(1)\top},d^{(2)\top})$, the action~\eqref{eq:fermion_action} reads
\be
S=\frac{1}{2}\int d^4x 
\Psi^\top
\begin{pmatrix}
P_1 & \im P_2 \\
-\im P_2^\top & P_3 \\
\end{pmatrix}
\Psi,
\label{eq:fermion_action_majorana_two}
\ee
where
\be
P_1=
\begin{pmatrix}
H_{uu}^{\rm AS}+h_{uu}^{\rm AS}+h_{uu}^{\prime\rm AS} & \im H_{uu}^{\rm S}+\im h_{uu}^{\rm AS}-\im h_{uu}^{\prime\rm AS}  \\
-\im H_{uu}^{\rm S}+\im h_{uu}^{\rm AS}-\im h_{uu}^{\prime\rm AS} & H_{uu}^{\rm AS}-h_{uu}^{\rm AS}-h_{uu}^{\prime\rm AS}  
\end{pmatrix} ,
\ee
\be
 P_2=
\begin{pmatrix}
-\im\left(H_{ud}^{\rm S}+h_{ud}^{\rm AS}+h_{ud}^{\prime\rm AS}\right) &  H_{ud}^{\rm AS}+ h_{ud}^{\rm AS}- h_{ud}^{\prime\rm AS}  \\
- H_{ud}^{\rm AS}+ h_{ud}^{\rm AS}- h_{ud}^{\prime\rm AS} & \im\left(H_{ud}^{\rm S}-h_{ud}^{\rm AS}-h_{ud}^{\prime\rm AS} \right) 
 \end{pmatrix} ,
\ee
\be
P_3=
\begin{pmatrix}
H_{dd}^{\rm AS}+h_{dd}^{\rm AS}+h_{dd}^{\prime\rm AS} & \im H_{dd}^{\rm S}+\im h_{dd}^{\rm AS}-\im h_{dd}^{\prime\rm AS}  \\
-\im H_{dd}^{\rm S}+\im h_{dd}^{\rm AS}-\im h_{dd}^{\prime\rm AS} & H_{dd}^{\rm AS}-h_{dd}^{\rm AS}-h_{dd}^{\prime\rm AS}  
\end{pmatrix} .
\ee
We change variables of the Grassmann integration as
\be
(u^{(1)},u^{(2)},d^{(1)},d^{(2)})\rightarrow(u^{(1)},u^{(2)},C\gamma_5d^{(1)},C\gamma_5d^{(2)}) ,
\ee 
and then the generating functional becomes
\be
Z 
= \pf
\begin{pmatrix}
P_1 & \im P_2 \\
-\im P_2^\top & P_3 
\end{pmatrix}
= \pf
\begin{pmatrix}
P_1 & \im \left(P_2 C^\top\gamma_5\right) \\
-\im\left(P_2 C^\top\gamma_5\right)^\top & C\gamma_5 P_3 C^\top\gamma_5 
\end{pmatrix}
.
\ee
The block matrix $P_2 C^\top\gamma_5$ must be semi-positive to satisfy the condition 1, which gives the constraints on the off-diagonal components of all the parameters in flavor space.
The condition $2$ is written as 
\bea
C\gamma_5 P_3 C^\top\gamma_5 &=&-P_{1}^\dagger,
\label{eq:positivity1}
\\
\left(P_2 C^\top\gamma_5\right)^\dagger &=& P_2 C^\top\gamma_5 .
\label{eq:positivity2}
\eea
From Eq.~\eqref{eq:positivity1}, we have
\bea
&&
\left(M\right)^\dagger_{uu}=\left(M\right)_{dd}, \
\left(M_5\right)^\dagger_{uu}=\left(M_5\right)_{dd}, \
\nonumber \\
-&&
\left(D_\mu\right)^\dagger_{uu}=\left(D_\mu\right)_{dd}, \
\left(D_{5\mu}\right)^\dagger_{uu}=\left(D_{5\mu}\right)_{dd}, \
-\left(\Sigma_{\mu\nu}\right)^\dagger_{uu}=\left(\Sigma_{\mu\nu}\right)_{dd}, \
\label{eq:fermion_matrix_majorna_sol1}
\\
&&
\left(m^{\rm S}\right)^\dagger_{uu}=\left(m^{\prime\rm S}\right)_{dd} , \
\left(m_5^{\rm S}\right)^\dagger_{uu}=\left(m_5^{\prime\rm S}\right)_{dd} , \
\nonumber \\
-&&
\left(d_\mu^{\rm AS}\right)^\dagger_{uu}=\left(d_\mu^{\prime\rm AS}\right)_{dd} , \
\left(d_{5\mu}^{\rm S}\right)^\dagger_{uu}=\left(d_{5\mu}^{\prime\rm S}\right)_{dd} , \
-\left(\sigma_{\mu\nu}^{\rm AS}\right)^\dagger_{uu}=\left(\sigma_{\mu\nu}^{\prime\rm AS}\right)_{dd} , \
\\
&&
\left(m^{\prime\rm S}\right)^\dagger_{uu}=\left(m^{\rm S}\right)_{dd} , \
\left(m_5^{\prime\rm S}\right)^\dagger_{uu}=\left(m_5^{\rm S}\right)_{dd} , \
\nonumber \\
-&&
\left(d_\mu^{\prime\rm AS}\right)^\dagger_{uu}=\left(d_\mu^{\rm AS}\right)_{dd} , \
\left(d_{5\mu}^{\prime\rm S}\right)^\dagger_{uu}=\left(d_{5\mu}^{\rm S}\right)_{dd} , \
-\left(\sigma_{\mu\nu}^{\prime\rm AS}\right)^\dagger_{uu}=\left(\sigma_{\mu\nu}^{\rm AS}\right)_{dd} .
\eea
From Eq.~\eqref{eq:positivity2}, we also have
\bea
-&&
\left(M\right)^\dagger_{ud}=\left(M\right)_{ud}, \
-\left(M_5\right)^\dagger_{ud}=\left(M_5\right)_{ud}, \
\nonumber \\
-&&
\left(M\right)^\dagger_{du}=\left(M\right)_{du}, \
-\left(M_5\right)^\dagger_{du}=\left(M_5\right)_{du}, \
\nonumber \\
&&
\left(D_\mu\right)^\dagger_{ud}=\left(D_\mu\right)_{ud}, \
-\left(D_{5\mu}\right)^\dagger_{ud}=\left(D_{5\mu}\right)_{ud}, \
\left(\Sigma_{\mu\nu}\right)^\dagger_{ud}=\left(\Sigma_{\mu\nu}\right)_{ud}, \
\nonumber \\
&&
\left(D_\mu\right)^\dagger_{du}=\left(D_\mu\right)_{du}, \
-\left(D_{5\mu}\right)^\dagger_{du}=\left(D_{5\mu}\right)_{du}, \
\left(\Sigma_{\mu\nu}\right)^\dagger_{du}=\left(\Sigma_{\mu\nu}\right)_{du}, \
\\
-&&
\left(m^{\rm S}\right)^\dagger_{ud}=\left(m^{\prime\rm S}\right)_{ud} , \
-\left(m_5^{\rm S}\right)^\dagger_{ud}=\left(m_5^{\prime\rm S}\right)_{ud}, \
\nonumber \\
&&
\left(d_\mu^{\rm AS}\right)^\dagger_{ud}=\left(d_\mu^{\prime\rm AS}\right)_{ud} , \
-\left(d_{5\mu}^{\rm S}\right)^\dagger_{ud}=\left(d_{5\mu}^{\prime\rm S}\right)_{ud}, \
\left(\sigma_{\mu\nu}^{\rm AS}\right)^\dagger_{ud}=\left(\sigma_{\mu\nu}^{\prime\rm AS}\right)_{ud} .
\label{eq:fermion_matrix_majorna_sol2}
\eea
Equations \eqref{eq:fermion_matrix_majorna_sol1}--\eqref{eq:fermion_matrix_majorna_sol2} and the semi-positivity of $P_2 C^\top\gamma_5$ guarantee the semi-positivity of the Pfaffian.
The result is shown in Table \ref{tab2}.

\begin{table}[p]
\centering
\begin{tabular}{|c|c|c|c|}
\hline
 term & condition1 & condition2 & examples
\\ \hline \hline
  $\bar{\psi}
 \begin{pmatrix}
 M & \widetilde{M} \\
\widetilde{M}^\prime &  M^\dagger \\
\end{pmatrix}
\psi$ & \checkmark & 
$\begin{matrix}
\widetilde{M}=-\widetilde{M}^\dagger
\\
\widetilde{M}^\prime=-\widetilde{M}^{\prime\dagger}
\end{matrix}$ 
& 
$\begin{matrix}
\text{degenerate Dirac mass}
\\
\text{Wilson term}
\end{matrix}$ 
\\ \hline
   $\bar{\psi}
\gamma_5
 \begin{pmatrix}
  M_{5} & \widetilde{M}_{5} \\
\widetilde{M}_{5}^\prime &   M_{5}^\dagger \\
\end{pmatrix}
\psi$ 
& \checkmark &
$\begin{matrix}
\widetilde{M}_5=-\widetilde{M}_5^\dagger
\\
\widetilde{M}_5^\prime=-\widetilde{M}_5^{\prime\dagger}
\end{matrix}$ 
& chirally twisted mass \cite{[{For a review, }]Frezzotti:2004pc}
\\ \hline
$\bar{\psi}\gamma_\mu
\begin{pmatrix}
  D_\mu & \widetilde{D}_\mu \\
  \widetilde{D}_\mu^{\prime} &  -D_\mu^\dagger
\end{pmatrix}
\psi$ 
& \checkmark & 
$\begin{matrix}
 \widetilde{D}_\mu=\widetilde{D}_\mu^\dagger \\
 \widetilde{D}_\mu^\prime=\widetilde{D}_\mu^{\prime\dagger} 
\end{matrix}$ 
& 
$\begin{matrix}
\text{gauge field}
\\
\text{isospin chemical potential \cite{Kogut:2002tm,*Kogut:2002zg,*Kogut:2004zg,Cea:2009ba,Cea:2012ev,Detmold:2012wc,Endrodi:2014lja}}
\\
\text{isospin electric field~\cite{Yamamoto:2012bd}}
\end{matrix}$ 
\\ \hline
$\bar{\psi}\gamma_\mu\gamma_5
\begin{pmatrix}
  D_{5\mu} & \widetilde{D}_{5\mu} \\
  \widetilde{D}^\prime_{5\mu} & D_{5\mu}^\dagger \\
\end{pmatrix}
\psi$
& \checkmark & 
$\begin{matrix}
 \widetilde{D}_{5\mu}=-\widetilde{D}_{5\mu}^\dagger \\
 \widetilde{D}_{5\mu}^\prime=-\widetilde{D}_{5\mu}^{\prime\dagger}
\end{matrix}$ 
&
$\begin{matrix}
\text{isospin axial gauge field}
\\
\text{chiral chemical potential \cite{Yamamoto:2011gk,*Yamamoto:2011ks,Buividovich:2013hza,Braguta:2015zta,Braguta:2015owi}}
\end{matrix}$ 
\\ \hline
$\bar{\psi}\gamma_\mu\gamma_\nu
\begin{pmatrix}
  \Sigma_{\mu\nu} & \widetilde{\Sigma}_{\mu\nu} \\
  \widetilde{\Sigma}_{\mu\nu}^{\prime} & -\Sigma_{\mu\nu}^\dagger \\
\end{pmatrix}
\psi$ 
&\checkmark& 
$\begin{matrix}
  \widetilde{\Sigma}_{\mu\nu} = \widetilde{\Sigma}_{\mu\nu}^\dagger
\\
  \widetilde{\Sigma}_{\mu\nu}^\prime = \widetilde{\Sigma}_{\mu\nu}^{\prime\dagger}  
\end{matrix}$
&
\\ \hline
 $\begin{matrix}
  \bar{\psi}_c
  \begin{pmatrix}
  m^{\rm S} & \widetilde{m}^{\rm S} \\
  \widetilde{m}^{\rm S} &  m^{\prime\rm S}
  \end{pmatrix}
  \psi
 \\
  +\bar{\psi}
  \begin{pmatrix}
  m^{\prime\rm S\dagger} & -\widetilde{m}^{\rm S\dagger}  \\
  -\widetilde{m}^{\rm S\dagger}  &  m^{\rm S\dagger}  \\
  \end{pmatrix}
  \psi_c
 \end{matrix}$
 & \checkmark &&
\\ \hline
$\begin{matrix}
 \bar{\psi}_c
 \begin{pmatrix}
  m_{5}^{\rm S} & \widetilde{m}_{5}^{\rm S} \\
  \widetilde{m}_{5}^{\rm S} &  m_{5}^{\prime\rm S} \\
 \end{pmatrix}
 \psi
\\
 + \bar{\psi}
 \begin{pmatrix}
   m_{5}^{\prime\rm S\dagger} & -\widetilde{m}_{5}^{\rm S\dagger}  \\
   -\widetilde{m}_{5}^{\rm S\dagger}  &  m_{5}^{\rm S\dagger}  \\
 \end{pmatrix}
 \psi_c 
 \end{matrix}$
&  \checkmark &&
\\ \hline
$\begin{matrix}
 \bar{\psi}_c\gamma_\mu
 \begin{pmatrix}
   d_{\mu}^{\rm AS} & \widetilde{d}_{\mu}^{\rm AS} \\
   -\widetilde{d}_{\mu}^{\rm AS} &  d_{\mu}^{\prime\rm AS} \\
 \end{pmatrix}
 \psi
\\
 +\bar{\psi}\gamma_\mu
 \begin{pmatrix}
   -d_{\mu}^{\prime\rm AS\dagger} & \widetilde{d}_{\mu}^{\rm AS\dagger}  \\
   -\widetilde{d}_{\mu}^{\rm AS\dagger}  &  -d_{\mu}^{\rm AS\dagger}  \\
 \end{pmatrix}
 \psi_c
\end{matrix}$
& \checkmark &&
\\ \hline
$\begin{matrix}
 \bar{\psi}_c\gamma_\mu\gamma_5
 \begin{pmatrix}
   d_{5\mu}^{\rm AS} & \widetilde{d}_{5\mu}^{\rm AS} \\
   -\widetilde{d}_{5\mu}^{\rm AS} &  d_{5\mu}^{\prime\rm AS} \\
 \end{pmatrix}
 \psi
\\
 + \bar{\psi}\gamma_\mu\gamma_5
 \begin{pmatrix}
   d_{5\mu}^{\prime\rm AS\dagger} & -\widetilde{d}_{5\mu}^{\rm AS\dagger}  \\
   \widetilde{d}_{5\mu}^{\rm AS\dagger}  & d_{5\mu}^{\rm AS\dagger}  \\
 \end{pmatrix}
 \psi_c 
\end{matrix}$
& \checkmark &&
\\ \hline
$\begin{matrix}
 \bar{\psi}_c\gamma_\mu\gamma_\nu
 \begin{pmatrix}
   \sigma_{\mu\nu}^{\rm AS} & \widetilde{\sigma}_{\mu\nu}^{\rm AS} \\
   -\widetilde{\sigma}_{\mu\nu}^{\rm AS} &  \sigma_{\mu\nu}^{\prime\rm AS} \\
 \end{pmatrix}
 \psi
\\
 + \bar{\psi}\gamma_\mu\gamma_\nu
 \begin{pmatrix}
   -\sigma_{\mu\nu}^{\prime\rm AS\dagger} & \widetilde{\sigma}_{\mu\nu}^{\rm AS\dagger}  \\
   -\widetilde{\sigma}_{\mu\nu}^{\rm AS\dagger}  &  -\sigma_{\mu\nu}^{\rm AS\dagger}  \\
 \end{pmatrix}
 \psi_c 
\end{matrix}$
& \checkmark &&
\\ \hline
\end{tabular}
\caption{\label{tab2}
Summary table of sign-problem-free terms of the two-flavor Dirac fermion.
In the column of condition 1, the checkmark stands for the constraint by the semi-positivity of $P_2 C^\top\gamma_5$ to the off diagonal components.
}
\end{table}

For two flavors, the sign-problem-free classes are enlarged.
A known example is the two-flavor QCD Dirac operator with the degenerate mass $M$ and nonzero isospin chemical potential $\mu_{\pi}$,
\be
D=
\begin{pmatrix}
\gamma_\mu D_\mu + M+\gamma^4 \mu_\pi & 0 \\
0 & \gamma_\mu D_\mu + M-\gamma^4 \mu_\pi
\end{pmatrix}
, \quad d = d' = 0 ,
\label{eq:dirac_iso}
\ee
with $D_\mu = \partial_\mu + \im A_\mu^a T^a = -D_\mu^\dagger$, and real $M$, $\mu_\pi \in \bf{R}$.
The Dirac operator~\eqref{eq:dirac_iso} satisfies Eqs.~\eqref{eq:fermion_matrix_majorna_sol1}--\eqref{eq:fermion_matrix_majorna_sol2} and the semi-positivity condition $P_2 C^\top\gamma_5=0$.
Thus the isospin chemical potential is sign-problem free unlike the baryon chemical potential.
Another example is the Wilson-Dirac operator.
The Wilson term $M$ satisfies Eq.~\eqref{eq:fermion_matrix_majorna_sol1}.
Thus the Wilson-Dirac operator is semi-positive for two flavors, while it is not for one flavor.

\section{miscellaneous}
\label{sec:mis}

Several comments are listed here:

\begin{itemize}

\item
The derived conditions are sufficient conditions, not necessary conditions, for sign-problem-free classes. 
Other sign-problem-free classes will be possible.
For example, the fermions obtained by the basis transformation from sign-problem-free fermions, such as spatially twisted chemical potential \cite{Fukuda:2013ada}, are also sign-problem free.

\item
The classification is independent of whether the parameter is a dynamical field or an external source.
When the parameter is dynamical, the integral is taken after the fermion integral.

\item
While all the components are sign-problem free for dynamical gauge fields, the corresponding physical situations depend on the components for external gauge fields.
An external magnetic field is sign-problem free but an external electric field has the sign problem \cite{[{For a review, }]D'Elia:2012tr}.
Similarly, an external axial electric field is sign-problem free but an external axial magnetic field has the sign problem. 

\item
For special internal groups, the classes of sign-problem-free terms are enlarged.
For example, baryon chemical potential can be sign-problem free for gauge groups in real or pseudo-real representation.
Such known examples are the fundamental representations of SU(2)
\cite{Hands:1999md,Kogut:2001na,Giudice:2004se,Cea:2006yd,Alles:2006ea,Chandrasekharan:2006tz,Hands:2006ve,*Hands:2010gd,Cea:2007vt,Lombardo:2008vc,Hands:2011ye,Cotter:2012mb,Boz:2013rca,Braguta:2016cpw} 
and G$_2$ \cite{Maas:2012wr,Wellegehausen:2013cya}, and the adjoint representation \cite{Hands:2000ei,Hands:2001ee}.

\item
The classification is applicable to the theory with four-fermion interactions.
A four-fermion action is converted to a bilinear-fermion action with an auxiliary field by the Hubbard-Stratonovich transformation.
When the resultant fermion action satisfies the Majorana positivity condition, the theory is sign-problem free.

\end{itemize}

\section{Summary} 
\label{sec:sum}

We discussed sign-problem-free relativistic fermions on the basis of the Majorana positivity.
The results are summarized in Tables~\ref{tab1} and~\ref{tab2}.
All known sign-problem-free terms were classified and some new sign-problem-free terms were found.
The results will be immediately applicable to the simulation of the Dirac fermion theory.
Although the computation of an indefinite Pfaffian is troublesome in general, the simulation of a semi-positive Pfaffian is the same as the standard simulation of a semi-positive determinant.
The classification by the Majorana positivity will be also effective for physical Majorana fermion systems.
The Majorana fermions are predicted in particle physics beyond the Standard Model, such as neutrinos and supersymmetry, and also focused on in condensed matter physics, such as quantum wires and topological superconductors \cite{[{For a review, }]Elliott:2014iha}.

\begin{acknowledgements}
T.~H.~was supported by JSPS Grant-in-Aid for Scientific Research (Grant No.~JP16J02240).
A.~Y.~was supported by JSPS KAKENHI (Grant No.~JP15K17624).   
\end{acknowledgements}

\appendix

\section{Derivation of the Majorana positivity condition}
\label{sec:appA}
We here prove the Majorana positivity condition.
From the condition 1, $P_2$ is written as 
\be
P_2=U^\dagger \sqrt{\Lambda} \sqrt{\Lambda} U ,
\ee
where $U$ is an unitary matrix, and $\Lambda$ is a diagonal matrix, whose components are nonnegative.
From the condition 2, $P$ is equivalently written by the form
\be
P=
\begin{pmatrix}
P_1 & \im P_2 \\
-\im P_2^* & P_1^* 
\end{pmatrix}
\label{eq:positive}
.
\ee
Then $P$ is rewritten as
 \be
P=
\begin{pmatrix}
\sqrt{\Lambda}U^\dagger & 0 \\
0 & \sqrt{\Lambda} U^T  
\end{pmatrix}
P^\prime
\begin{pmatrix}
\sqrt{\Lambda}U^* & 0 \\
0 & \sqrt{\Lambda} U 
\end{pmatrix}
,
\ee
where 
\be
P^\prime=
\begin{pmatrix}
Q_N & \im \bm 1_N  \\
-\im \bm 1_N & Q_N^*
\end{pmatrix} ,
\ee
with $Q_N=[\sqrt{\Lambda}U^\dagger]^{-1}P_1[\sqrt{\Lambda}U^*]^{-1}$, and $\bm 1_N$ is the $N\times N$ unit matrix.
We have 
 \be
\pf(P)=
\pf(P^\prime)
{\rm det}
\begin{pmatrix}
\sqrt{\Lambda}U^* & 0 \\
0 & \sqrt{\Lambda} U 
\end{pmatrix}
,
\ee
and the determinant is semi-definite, so that the our goal is to show the semi-positivity of $\pf(P^\prime)$. 
We expand $\pf(P^\prime)$ by the number $r$ to count how many times the off-diagonal component $\im \bm 1_N$ contributes.
For a certain value of $r$, the contribution reads
\be
\sum_{\{m_1,\ldots,m_r\}}
\pf (Q_{N-r})
\pf
\begin{pmatrix}
0 & \im \bm 1_r \\
-\im \bm 1_r & 0 
\end{pmatrix}
\pf (Q_{N-r}^*)
,
\ee
where $Q_{N-r}$ is the $(N-r)\times (N-r)$ matrix obtained from $Q_N$ by removing its $m_1$-, $\ldots,m_r$-th rows and columns, 
and $\sum_{\{m_1,\ldots,m_r\}}$ denotes the summation over all possible sets $\{m_1,\ldots,m_r\}$ satisfying $1\leq m_1<\ldots<m_r\leq N$. 
Then $\pf(P^\prime)$ is given as
\bea
 \pf (P^\prime)
&=&\sum_{r=0}^N \sum_{\{m_1,\ldots,m_r\}} \pf (Q_{N-r})
\pf
\begin{pmatrix}
0 & \im \bm 1_r  \\
-\im \bm 1_r  & 0 
\end{pmatrix}
\pf (Q_{N-r}^*)
\nonumber
\\
&=&\sum_{r=0}^N\sum_{\{m_1,\ldots,m_r\}}|\pf (Q_{N-r})|^2
(-)^{\frac{r(r-1)}{2}}\im^r 
\nonumber
\\
&=&\sum_{r:\rm even}\sum_{\{m_1,\ldots,m_r\}}|\pf (Q_{N-r})|^2 
+\im\sum_{r:\rm odd}\sum_{\{m_1,\ldots,m_r\}}|\pf (Q_{N-r})|^2 .
\label{eq:pf}
\eea
The Pfaffian of an odd-dimensional anti-symmetric matrix is zero by definition.
When $N$ is even, the second term in the last line in Eq.~\eqref{eq:pf} vanishes since $(N-r)$ is odd.
Therefore we can show $\pf (P^\prime) \ge 0$ and then $\pf (P) \ge 0$.
(On the other hand  when $N$ is odd,  the first term in the last line in Eq.~\eqref{eq:pf} vanishes.
This, in turn, shows that $\pf (P)$ is pure imaginary.
Although the generating functional is not semi-positive, it can be semi-positive and sign-problem free by trivial phase rotation $Z \to -\im Z \ge 0$.)

\section{Gamma matrix}
\label{sec:appB}

We use the Euclidean gamma matrices in the chiral basis
\be
\gamma_i=
\begin{pmatrix}
0 & \im \sigma_i \\
-\im \sigma_i & 0
\end{pmatrix} ,
\quad
\gamma_4=
\begin{pmatrix}
0 & 1 \\
1 & 0
\end{pmatrix} ,
\quad
\gamma_5=
\begin{pmatrix}
-1 & 0 \\
0 & 1
\end{pmatrix} ,
\ee
and 
\bea
\gamma_i\gamma_j &=&
\begin{pmatrix}
\sigma_i\sigma_j & 0 \\
0 & \sigma_i\sigma_j
\end{pmatrix} ,
\quad
\gamma_i\gamma_4 =
\begin{pmatrix}
\im\sigma_i & 0 \\
0 & -\im\sigma_i
\end{pmatrix} ,
\nonumber
\\
\gamma_i\gamma_5 &=&
\begin{pmatrix}
0 & \im\sigma_i \\
\im\sigma_i & 0
\end{pmatrix} ,
\quad
\gamma_4\gamma_5 =
\begin{pmatrix}
0 & 1 \\
-1 & 0
\end{pmatrix} .
\eea
The charge conjugation matrix is
\be
C = \gamma_2\gamma_4 =
\begin{pmatrix}
\im\sigma_2 & 0 \\
0 & -\im\sigma_2
\end{pmatrix} ,
\ee
satsfying
\be
C^*=C, \;\;\;\; C^\top=-C, \;\;\;\; C^\dagger C=1, \;\;\;\; C^2=-1 .
\ee
We also define
\begin{equation}
C_L =
\begin{pmatrix}
1 & 0 \\
0 & -\im \sigma_2
\end{pmatrix}
.
\end{equation}
The matrix elements of Eq.~\eqref{eq:oneh1} are
\bea
H_{RR} &=& \im\sigma_2( -\im\sigma_i D_i + D_4 + \im\sigma_i D_{5i} - D_{54} ) ,
\\
H_{RL} &=& \sigma_2 (M+M_5+ \sigma_i \sigma_j \Sigma_{ij} - 2\im\sigma_i \Sigma_{i4} ) \sigma_2 ,
\\
H_{LR} &=& M-M_5 + \sigma_i \sigma_j \Sigma_{ij} + 2\im\sigma_i \Sigma_{i4} ,
\\
H_{LL} &=& -\im(\im\sigma_i D_i + D_4 + \im\sigma_i D_{5i} + D_{54} ) \sigma_2 ,
\eea
those of Eq.~\eqref{eq:oneh2} are
\bea
h_{RR} &=& \im\sigma_2(m-m_5 + \sigma_i \sigma_j \sigma_{ij} + 2\im\sigma_i \sigma_{i4}) ,
\\
h_{RL} &=& \sigma_2 (\im\sigma_i d_i + d_4 + \im\sigma_i d_{5i} + d_{54}) \sigma_2 ,
\\
h_{LR} &=& -\im\sigma_i d_i + d_4 + \im\sigma_i d_{5i} - d_{54} ,
\\
h_{LL} &=& -\im(m+m_5+ \sigma_i \sigma_j \sigma_{ij} - 2\im\sigma_i \sigma_{i4} )\sigma_2 ,
\eea
and those of Eq.~\eqref{eq:oneh3} are
\bea
h_{RR}' &=& -\im\sigma_2 (m'+m'_5+ \sigma_i \sigma_j \sigma'_{ij} - 2\im\sigma_i \sigma'_{i4} ) ,
\\
h_{RL}' &=& -\sigma_2 ( -\im\sigma_i d'_i + d'_4 + \im\sigma_i d'_{5i} - d'_{54} ) \sigma_2 ,
\\
h_{LR}' &=& -(\im\sigma_i d'_i + d'_4 + \im\sigma_i d'_{5i} + d'_{54}) ,
\\
h_{LL}' &=& \im(m'-m'_5 + \sigma_i \sigma_j \sigma'_{ij} + 2\im\sigma_i \sigma'_{i4}) \sigma_2 .
\eea


\bibliography{majorana}

\end{document}